\documentclass[conference,a4paper]{IEEEtran}

\usepackage{cite}
\usepackage{graphicx,color,epsfig,rotating}
\usepackage{amsfonts,amsmath,amssymb,bbm}
\usepackage{algorithm}
\usepackage{algpseudocode}
\usepackage{amsmath}
\usepackage{cite}
\usepackage{mdwtab} 
\usepackage{placeins}
\usepackage{psfrag, graphicx}
\usepackage[latin1]{inputenc}
\usepackage{amssymb}
\usepackage{makeidx}
\usepackage{epstopdf}
\usepackage{enumitem}

\usepackage{subcaption}
\usepackage{caption}

\long\def\comment#1{}

\newfont{\bbb}{msbm10 scaled 700}

\newfont{\bb}{msbm10 scaled 1100}

\newcommand{\bv}{{\bf b}}

\newcommand{\sv}{{\bf s}}

\newcommand{\Mm}{{\bf M}}

\newcommand{\Xm}{{\bf X}}

\newcommand{\Ac}{{\cal A}}

\newcommand{\Mc}{{\cal M}}
\newcommand{\Nc}{{\cal N}}

\newcommand{\Pc}{{\cal P}}

\newcommand{\Sc}{{\cal S}}
\newcommand{\Tc}{{\cal T}}

\newcommand{\Zc}{{\cal Z}}

\newcommand{\muv}{\hbox{\boldmath$\mu$}}

\newcommand{\eqdef}{\stackrel{\Delta}{=}}

\newcommand{\be}{\begin{equation}}
\newcommand{\ee}{\end{equation}}
\newcommand{\bea}{\begin{eqnarray}}
\newcommand{\eea}{\end{eqnarray}}

\newtheorem{defn}{Definition}
\newtheorem{remark}{Remark}

\begin{document}



\title{A New Design Framework for Heterogeneous Uncoded Storage Elastic Computing}

\author{Mingyue Ji$^{1}$, Xiang Zhang$^{1}$ and Kai Wan$^{2}$
\thanks{The authors are with the Department of Electrical Engineering,
University of Utah, Salt Lake City, UT 84112, USA. (e-mail: nicholas.woolsey@utah.edu, rchen@ece.utah.edu and mingyue.ji@utah.edu)}
}

\author{
    \IEEEauthorblockN{ Mingyue Ji$^{1}$, Xiang Zhang$^{1}$,
		and Kai Wan$^{2}$ }
	\IEEEauthorblockA{$^1$University of Utah, \;\; $^2$Technische Universit\"at Berlin\\
		Email: \{mingyue.ji@utah.edu, xiang.zhang@utah.edu, kai.wan@tu-berlin.de\}}

}

\maketitle

\thispagestyle{empty}
\pagestyle{empty}


\begin{abstract}
Elasticity is one important feature in modern cloud computing systems and can result in computation failure or significantly increase computing time. Such elasticity means that virtual machines over the cloud can be preempted under a short notice (e.g., hours or minutes) if a high-priority job appears; on the other hand, new virtual machines may become available over time to compensate the computing resources. Coded Storage Elastic Computing (CSEC) introduced by Yang et al. in 2018 is an effective and efficient approach to overcome the elasticity and it costs relatively less storage and computation load. However, one of the limitations of the CSEC is that it  may only be applied to certain types of computations (e.g., linear) and may be challenging to be applied to more involved computations because the coded data storage and approximation are often needed. Hence, it may be preferred to use uncoded storage by directly copying data into the virtual machines. In addition, based on our own measurement, virtual machines on Amazon EC2 clusters often have 
heterogeneous computation speed even if they have exactly the same configurations (e.g., CPU, RAM, I/O cost). In this paper, we introduce a new optimization framework on Uncoded Storage Elastic Computing (USEC) systems with heterogeneous computing speed to minimize the overall computation time. Under this framework, we propose optimal solutions of USEC systems with or without straggler tolerance using different storage placements. Our proposed algorithms are evaluated using power iteration applications on Amazon EC2. 
\end{abstract}



\section{Introduction}
\label{section: intro}

Coded Storage Elastic Computing (CSEC) system introduced by Yang et al. in \cite{yang2018coded} is an effective approach to overcome the elasticity of modern cloud computing system, where elasticity means that Virtual Machines (VMs) on the cloud systems, e.g., instances on Amazon EC2,  can be preempted under a short notice (e.g., hours or minutes) if a high-priority job appears; on the other hand, new VMs may become available over time to compensate the computing resources. Such elasticity can result in computation failure or significantly increase computing time. In \cite{yang2018coded}, using a Maximum Distance Separable (MDS) coded storage placement, the authors proposed a {\em cyclic} computation assignment scheme such that no redundant computation is needed when the number of available VMs $N_t$ is between $L$ and $N$ where $N$ is the maximum number of VMs in the systems and $L$ is the smallest number of VMs in the system. In \cite{dau2020optimizing}, the authors introduced a new metric, called transition waste, which is defined as the difference between the total number of changes and the number of necessary changes of the computation assignment if some VMs become preempted during one computation or time step. This problem is combinatorial and is challenging to be solved in general. The authors proposed new algorithms using {\em shifted cyclic} task allocation to reduce the transition waste and showed it is optimal under some parameter settings. In \cite{kiani2021cec}, the authors proposed two hierarchical schemes that can further speed up the USEC system by effectively allocating tasks among available nodes while the encoding and decoding complexity may be increased. Some important limitations of \cite{yang2018coded,dau2020optimizing,kiani2021cec} include the assumption that all available VMs have the same computing speed or the proposed schemes do not consider the heterogeneous computing speed among machines, and all VMs have the homogeneous storage constraint. In practice, based on our own measurement \cite{woolsey2021practicalcec}, the computing speed among VMs can be significantly different even if they have exactly the same configurations, e.g., same CUP, RAM and I/O cost. In \cite{woolsey2020heterogeneous}, the authors considered the elastic computing systems with heterogeneous computing speed and homogeneous storage constraint, and formulated a new CSEC framework, that is to minimize the overall computation time, using a combinatorial optimization approach. In addition, one exact optimal solution is provided and can be achieved using the {\em filling algorithm}, which is a low-complexity iterative algorithm that can complete within $N_t$ iterations, where $N_t$ is the number of available VMs at time step $t$. Later, in \cite{woolsey2021cec}, the authors considered the CSEC system with both heterogeneous computing speed and heterogeneous storage constraint, and formulated a new combinatorial optimization framework based on the result in 
\cite{woolsey2020heterogeneous} and designed algorithms to achieve the optimal computation time. Under the assumption of heterogeneous computing speed, in \cite{woolsey2021practicalcec}, the authors made preliminary attempts to study the scenario where both elasticity and stragglers are present and proposed new algorithms using the idea of the filling algorithm.\footnote{Stragglers are often referred to as the machines with abnormally slower speed.} 
An achievable trade-off between computation time and straggler tolerance was established. In addition, the authors in \cite{woolsey2021practicalcec} implemented the proposed algorithms for heterogeneous CSEC systems using real applications on Amazon EC2 and demonstrated that large gain in terms of the computation time can be achieved by the proposed algorithms. 

Despite clear advantages of the CSEC systems such as less storage overhead,  it can only be applied to certain types of computations (e.g., linear) and may be challenging to be applied to more involved computations (e.g., deep learning) due to the coded data storage. In this case, approximation is often needed. Hence, it may be preferred to use uncoded storage by just copying the data into the virtual machines since computations can be operated directly over the original data in this case. 
We refer to such systems as Uncoded Storage Elastic Computing (USEC) systems. In this paper, we introduce a new optimization framework on USEC with heterogeneous computing speed to minimize the overall computation time. We propose solutions to USEC systems with or without straggler tolerance using different storage placements.

Our contributions are summarized as follows:
\begin{enumerate}
  \item When there is no straggler tolerance requirement, given the storage placement and the heterogeneous computing speed of VMs, we formulate a new USEC framework as a convex optimization problem which can be solved using typical convex optimization solvers. Further, we investigate the performance in terms of computation time using different uncoded storage placements. 
  \item We incorporate straggler tolerance into the above problem formulation and formulate it as a combinatorial optimization problem. In addition, we design a low-complexity algorithm to 
  achieve the optimal solution of the proposed optimization problem given the uncoded storage placement.
  \item We perform experiments using the proposed USEC framework with heterogeneous computing speed, and using 
 the power iteration application under a simple setup. We demonstrate that about $20\%$ gain in terms of computation time can be achieved using the proposed algorithms by taking the advantage of heterogeneous computing speed. 
\end{enumerate}

\paragraph*{Notation Convention}
We use $|\cdot|$ to represent the cardinality of a set or the length of a vector
and $[n] \eqdef \{1,2,\ldots,n\}$. 
A bold symbol such as $\boldsymbol{a}$ indicates a vector and $a[i]$ denotes the $i$-th element of $\boldsymbol{a}$. 
Calligraphic symbols such as $\Ac$ presents a set with numbers as its elements. Bold calligraphic symbols such as $\boldsymbol{\Ac}$ represents a set whose elements are sets (e.g., $\Ac$). 

\section{Network Model and Problem Formulation}
\label{sec: Network Model and Problem Formulation}

We consider a set of $N$ VMs jointly store an uncoded data matrix $\Xm$ with dimension $q \times r$, which 
 is row-wise partitioned in $\Xm = [\Xm_1; \Xm_2; \cdots; \Xm_G]$. With a slight abuse of notation, $\Xm_g, g \in [G]$ denotes both the {\em row sets} and {\em sub-matrices} of $\Xm$. In particular, the number of rows in each $\Xm_g, g\in [G]$ is $q/G$ and we index them as $[q/G]$. Each $\Xm_g$ is placed into $J$ machines. Let $\Nc_g = \{n: \Xm_g \in \Zc_n\}$ denote the set of VMs that stores $\Xm_g$ and $\mathcal{Z}_n$ be the storage placement for machine $n$. The set of the storage placements for all VMs is denoted by $\boldsymbol{\mathcal{Z}} = \{\mathcal{Z}_n, n \in [N]\}$. 
 Similar to \cite{yang2018coded}, the machines collectively perform matrix-vector computations over multiple computation steps. In a given time step only a subset of the $N$ machines are available to perform matrix computations. More specifically, in computation step $t$, a set of available machines  $\mathcal{N}_t \subseteq [N]$ with $|\Nc_t| = N_t$ aims to compute
\be
\boldsymbol{y}_t = \boldsymbol{X}\boldsymbol{w}_t,
\ee
where $\boldsymbol{w}_t$ is some vector of length $r$. The machines of $[N]\setminus \mathcal{N}_t$ are preempted. 

The VMs in $\mathcal{N}_t$ do not compute $\boldsymbol{y}_t$ directly. Instead, each machine $n \in \mathcal{N}_t$ computes $\Xm_{\Sc_n} \boldsymbol{w}_t$, where $\Sc_n \subset \Xm_g, \Xm_g \in \Zc_n$ denotes a row set in the sub-matrix $\Xm_g \in \Zc_n$. 
Then the results from VMs will be sent to the master machine to obtain $\boldsymbol{y}_t$.  
Let $\mathcal{T}_{g, n}$ denote the row set of sub-matrix $\Xm_g$ computed at machine $n \in \Nc_t$. 
\begin{defn} {\bf (Computation load)} 
Let the computation load matrix be $\boldsymbol{M}$ and each entry of $\boldsymbol{M}$, $[\boldsymbol{M}]_{g,n} = \mu[g,n]$, is the computation load of sub-matrix $\Xm_g$ at machine $n$ defined as  
\be \label{eq: compload_matrix}
\mu[g,n] \eqdef \frac{|\mathcal{T}_{g, n}|}{q/G}. 
\ee 
If $\Xm_g \notin \Zc_n$, $\mu[g,n]=0$. 
The computation load vector for $N$ machines, $\boldsymbol{\mu} = [\mu[1], \cdots, \mu[n]]$, is defined as 
\be \label{eq: compload_vector}
\mu[n] = \sum_{g \in [G]}\mu[g,n], \;\; \forall n \in \mathcal{N}_t, 
\ee
which is the sum of the fractions of rows of the corresponding stored sub-matrices computed by machine $n$ at time step $t$. \hfill $\Diamond$ 
\end{defn}
Note that $\mathcal{T}_{g, n}$, $\boldsymbol{M}$ and $\boldsymbol{\mu}$ 
may change with each time step, but reference to $t$ is omitted for ease of disposition. Moreover, the machines have varying computation speed defined by the strictly positive vector, $\boldsymbol{s}$, which is known for each time step and defined as follows. 
\begin{defn}
{\bf (Computation Speed)} The computation speed vector $\boldsymbol{s}$ is a length-$N$ vector with elements $s[n]$, $n\in[N]$, where $s[n]$ is the speed of machine $n$ measured as the inverse of the time it takes machine $n$ to compute all rows of one of its assigned sub-matrix. 
\hfill $\Diamond$
\end{defn}


The computation time is dictated by the VM that takes the most time to perform its assigned computations, and defined as follows. 
\begin{defn}
{\bf (Computation Time)} The computation time in a particular time step is defined as 
\be
\label{eq: comptime} 
c(\boldsymbol{M}) = c(\boldsymbol{\mu}) \eqdef \max_{n \in \Nc_t} \frac{\mu[n]}{s[n]} = \max_{n \in \Nc_t} \frac{\sum_{g \in [G]}\mu[g,n]}{s[n]}. 
\ee  
\hfill $\Diamond$
\end{defn}
\subsection{USEC without straggler tolerance}
We first formulate the optimization framework for the USEC systems without straggler tolerance. For a fixed storage placement $\boldsymbol{\mathcal{Z}}$, we can formulate the following optimization problem. 
\begin{subequations} \label{eq: uncoded opt}
\begin{align}
\underset{{\mathcal{T}}_{g,n} }{\text{minimize}} & \quad c\left(\boldsymbol{M}\right) \label{eq: uncoded 1} \\
\text{subject to:}  &\bigcup_{n \in \Nc_t: \Xm_g \in \Zc_n} \mathcal{T}_{g,n} = \left[\frac{q}{G}\right], \forall g \in [G]\label{eq: unocded constraint 1}. 
\end{align}
\end{subequations} 
It can be shown that the optimization problem (\ref{eq: uncoded opt}) is  equivalent to the following convex optimization problem.
\begin{subequations} \label{eq: uncoded opt 2}
\begin{align}
\underset{{\boldsymbol{M}} }{\text{minimize}} & \quad c\left(\boldsymbol{M}\right) = \max_{n \in  \Nc_t}  \frac{\sum_{g \in [G]}\mu[g,n]}{s[n]} \label{eq: uncoded 1} \\
\text{subject to:}  & \quad \sum_{n \in  \Nc_t: \Xm_g \in \Zc_n} \mu[g,n] = 1, \forall g \in [G], \label{eq: usec constraint mu 2}\\ 
& \quad \mu[g,n] = 0, \forall \Xm_g \notin \Zc_n, n \in \Nc_t, \\
& \quad 0 \leq \mu[g,n] \leq 1, \forall n \in \Nc_t.
\end{align}
\end{subequations}
It can be seen that by solving (\ref{eq: uncoded opt 2}), we can obtain the optimal computation assignment $\boldsymbol{M}^{\star}$, which can be used to find the corresponding $\mathcal{T}_{g,n}$ straightforwardly since each row in $\Xm_g$ is computed only once (see Section~\ref{sec: example} for examples).  

\subsection{USEC with straggler tolerance}

When straggler tolerance is incorporated into the USEC framework, we use the {\em redundant task assignment} approach, 
meaning that each row in $\Xm$ can be computed $1+S$ times in order to tolerate at most $S$ stragglers. 
This implies that the computation can be recovered when any $S$ machines, denoted by $\Sc$, of the available machines $\Nc_t$ become stragglers and $\Sc$ is not known a priori. 
Hence, this problem becomes a combinatorial optimization problem. In particular, 
a computation assignment within $\Xm_g$ is defined by $F_g$ disjoint sets of rows in $\Xm_g$, i.e., $\boldsymbol{\Mc}_g  = \{\mathcal{M}_{g,1},\ldots,\mathcal{M}_{g,F_g}\}$ such that $\bigcup_{f \in [F_g]} \mathcal{M}_{g,f} = \left[ \frac{q}{G}\right]$. 
Then, $F_g$ sets of machines, $\boldsymbol{\mathcal{P}}_g = \{\mathcal{P}_{g,1} , \ldots , \mathcal{P}_{g,F_g}\}$, which store and perform computation over $\Xm_g$, are defined such that $\mathcal{P}_{g,f} \subseteq \{n \in \Nc_t: \Xm_g \in \Zc_n\}$, $|\mathcal{P}_{g,f}|=1+S, \forall f\in [F_g]$ and machines in $\mathcal{P}_{g,f}$ computes the row set $\mathcal{M}_{g,f}$ in $\Xm_g$. 
Note that $\Tc_{g,n} = \bigcup_{f \in [F_g]: n \in \Pc_{g,f}} \Mc_{g,f}$. 
The sets 
$\boldsymbol{\Mc}_g$, $\boldsymbol{\Pc}_g$ 
and $F_g$ may vary with each time step based on  machines' availability.

In a given time step $t$, our goal is to design the task assignments, $\boldsymbol{\Mc}_g, \boldsymbol{\Pc}_g, g \in [G]$, 
such that the computation $\boldsymbol{y}_t = \boldsymbol{X}\boldsymbol{w}_t$ can be recovered when some VMs are stragglers that do not provide their assigned computations to the master machine. 

Then, we aim to design the computation assignment that minimizes the computation time of (\ref{eq: comptime}) resulting from the computation load matrix defined in (\ref{eq: compload_matrix}).
In time step $t$, given $\boldsymbol{\Zc}$, $\mathcal{N}_t$ and $\boldsymbol{s}$, the optimal computation time, $c^\star$, is the minimum of computation times defined by all possible task assignments,  
such that $S$ stragglers can be tolerated and the computation can be recovered. In particular $c^\star$ is the optimal value of the following combinatorial optimization problem. 
\begin{subequations} 
\label{eq: opt}
\begin{align}
&\underset{\boldsymbol{\Mc}_{g}, \boldsymbol{\Pc}_{g}}{\text{minimize}} \;\; c\left(\boldsymbol{M}\right) \\
&\text{s.t. } 
\bigcup_{f \in [F_g]} \mathcal{M}_{g,f} = \left[ \frac{q}{G}\right], \forall g \in [G],\label{eq: optprob_assign}\\
&\;\;\;\;\;\; |\mathcal{P}_{g, f}\setminus \Sc | \geq 1,\forall g \in [G], \mathcal{P}_{g, f} \in \boldsymbol{\mathcal{P}}_g, \forall \Sc \subset \mathcal{N}_t,|\Sc|=S. 
\end{align}
\end{subequations}
The optimization problem (\ref{eq: opt}) is combinatorial and the optimal solution is challenging. In the following, we will propose a novel low-complexity algorithm to achieve the optimal solution for this combinatorial optimization problem. Interestingly, the {\em filling algorithm} introduced in the 
CSEC framework with heterogeneous computing speed \cite{woolsey2021cec} or the heterogeneous storage-constrained private information retrieval problem \cite{woolsey2021pir} can be applied here to obtain the proposed optimal solution for (\ref{eq: opt}). 


\begin{figure}
\begin{subfigure}{1\columnwidth}
  \centering
  \includegraphics[width=0.65\columnwidth]{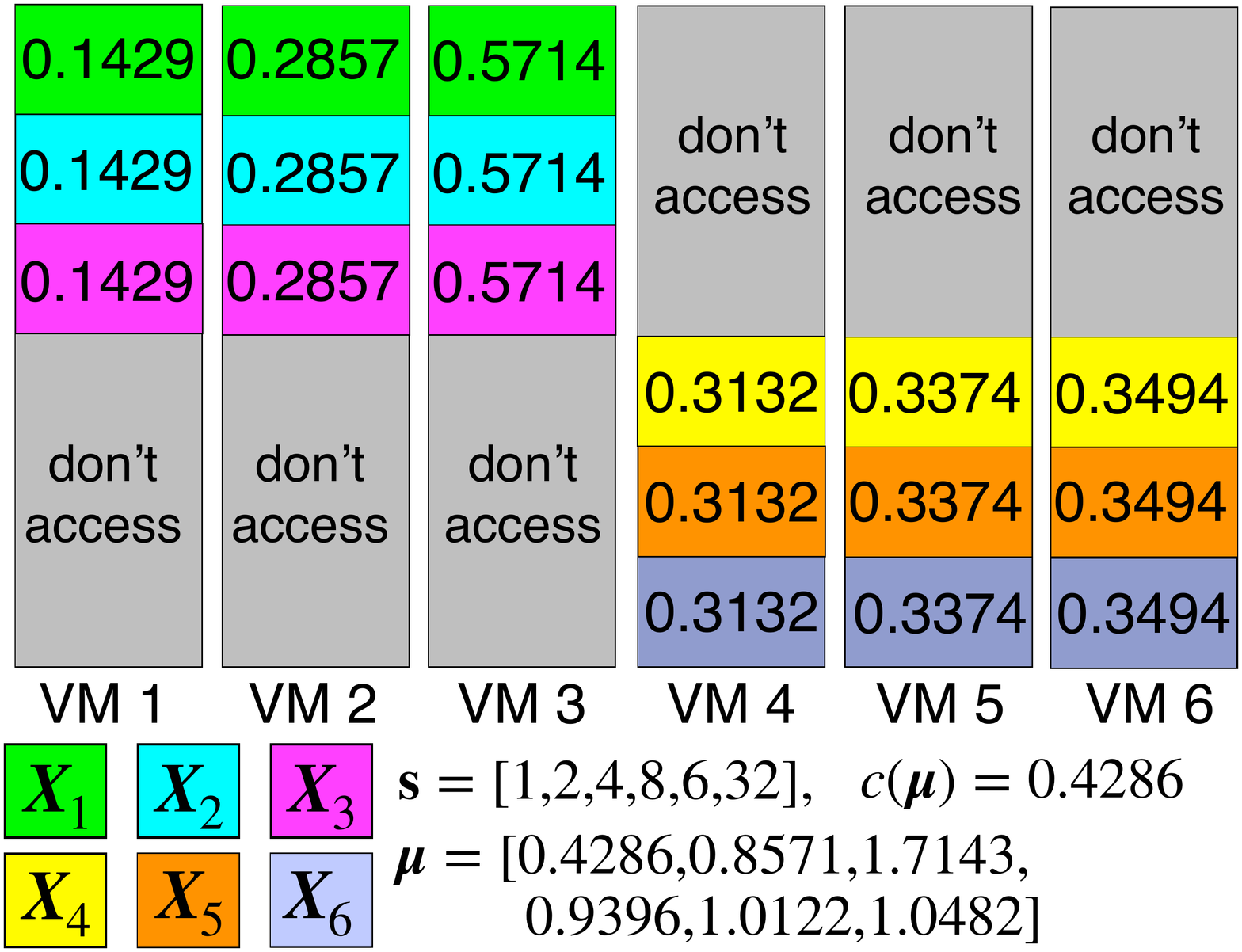}  
  \caption{Repetition placement.}
  \label{fig: Uncoded_Elastic_Computing_Repeition}
\end{subfigure}
\begin{subfigure}{1\columnwidth}
  \centering
  \includegraphics[width=0.65\columnwidth]{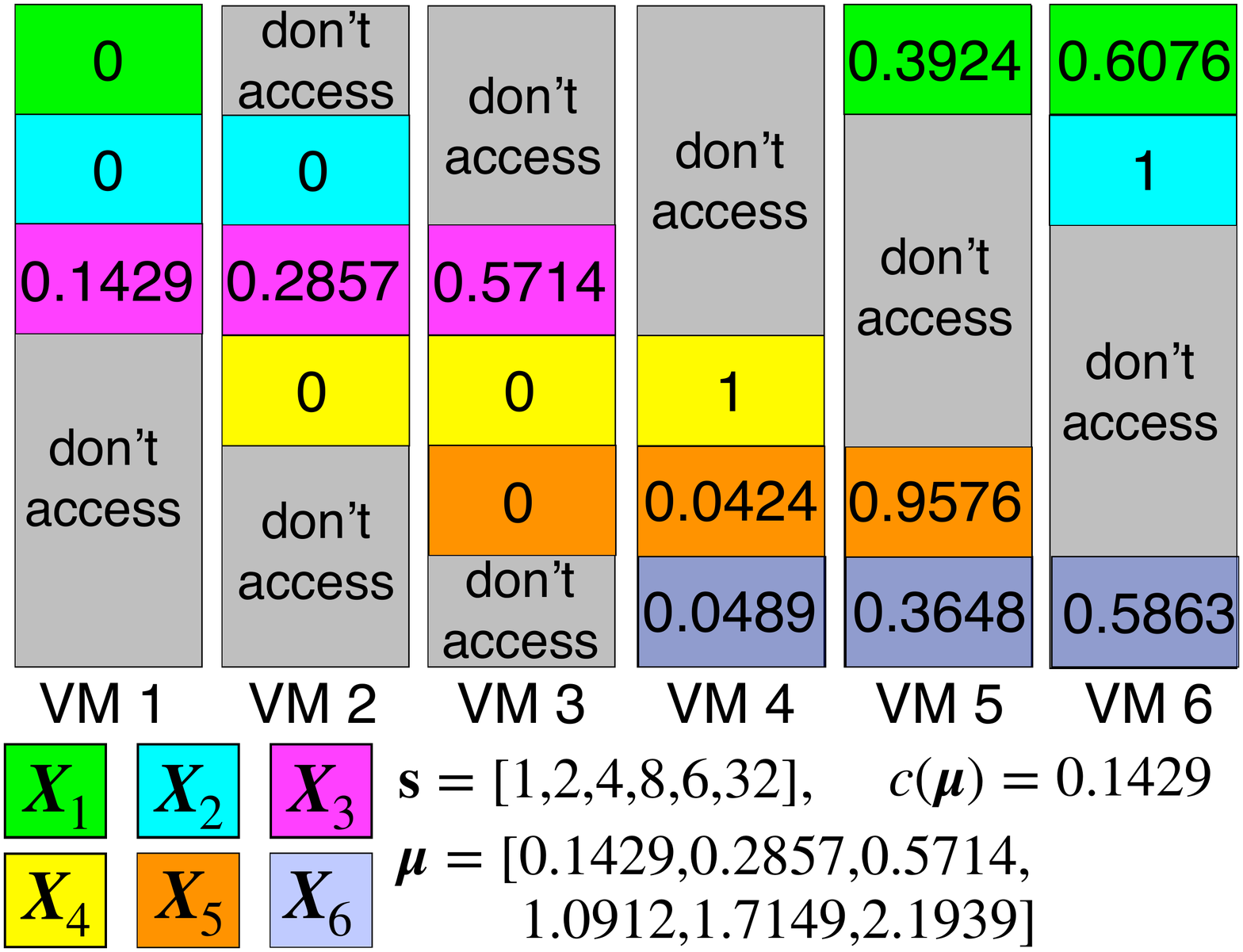} 
  \caption{Cyclic placement.}
  \label{fig: Uncoded_Elastic_Computing_Cyclic}
\end{subfigure} 
\caption{Illustration of the proposed USEC framework.} \label{fig: usec}
\end{figure} 
\begin{figure}
\centering
\includegraphics[width=0.75\columnwidth]{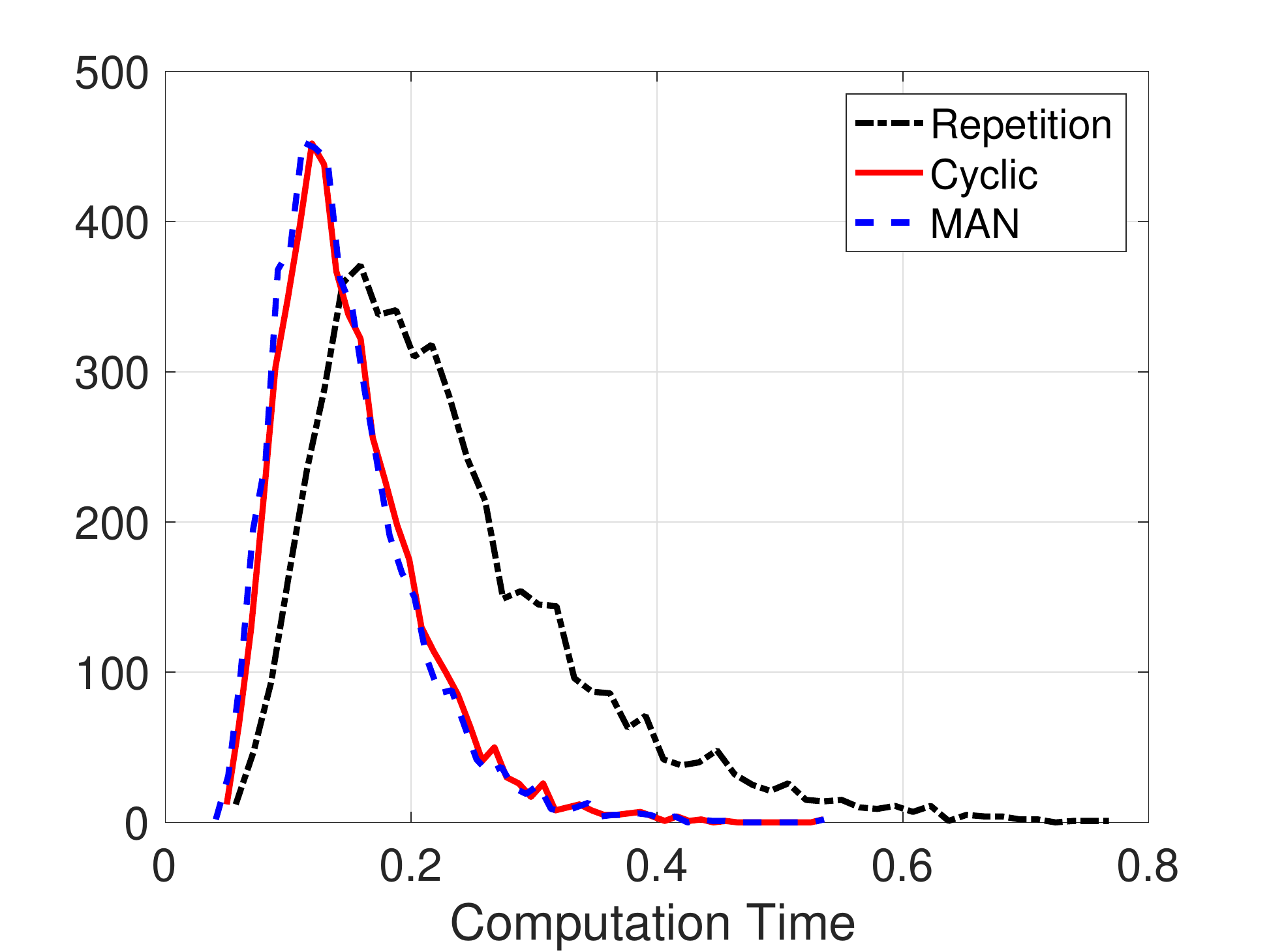}
\caption{Comparison of histograms of $C(\boldsymbol{M})$ for repetition, cyclic and MAN storage placements over 5000 realizations of the computing speed vector.}
\label{fig: uncoded_hist}
\end{figure}

\section{Examples} 
\label{sec: example}
In this section, we will illustrate two examples of the proposed USEC framework with and without straggler tolerance, respectively, under the homogeneous storage constraint. 
We consider two commonly used uncoded storage schemes, which are {\em fractional repetition placement} (referred to as repetition placement hereafter) and {\em cyclic placement}, which are widely used in the distributed storage and gradient coding literatures \cite{tandon2017gradient,ye2018communication,netanel2020tit}. In particular, 
we consider a USEC system with $N = 6$ VMs and the speed vector is $\sv = [1, 2, 4, 8, 16, 32]$. 
The data matrix $\Xm$ is partitioned into $G=6$ sub-matrices, each placed into $J=3$ machines. Fig.~\ref{fig:  usec} shows this system with repetition placement (Fig.\ref{fig: Uncoded_Elastic_Computing_Repeition}) and with cyclic placement (Fig.~\ref{fig: Uncoded_Elastic_Computing_Cyclic}), respectively. Let $N = N_t$, all $\mu[g,n], g \in [6], n \in [N]$ are computed by solving the convex optimization problem (\ref{eq: uncoded opt 2}). 
In Fig.~\ref{fig:  usec}, the colors represent the storage placement of each sub-matrix and the numbers inside represent the corresponding $\mu[g,n]$ for sub-matrix $g$ and machine $n$. 
The computation time for the cyclic placement is $c(\muv) = 0.1429$, which is significantly better than that of the repetition placement $c(\muv) = 0.4286$. However, interestingly, the cyclic placement is not necessarily better than the repetition placement for any speed vector. For example, if machines 3 and 4 are much faster than other VMs, then the repetition placement can be better than the cyclic placement since machines 3 and 4 stores the entire data matrix under the repetition placement. 
In order to have a better understanding of this phenomenon, we ran an experiment by randomly generating $\sv$ based on an exponential distribution. By solving the minimum computation time for each $\sv$ using (\ref{eq: uncoded opt}), we obtain the distribution of the computation time for these two storage placements shown in Fig.~\ref{fig: uncoded_hist}, where 
the cyclic placement (red) is much better than the repetition placement (yellow) in most realizations. In particular, there are only $68$ cyclic placement realizations out of $5000$ worse than repetition placement realizations. Although these results show the promising performance of the cyclic storage placement, it is not optimal in general. For example, using the Maddah-Ali Niesen coded caching (MAN) storage placement scheme \cite{maddah2014fundamental} to repeat the same experiment, 
we can obtain slightly better results as shown in Fig.~\ref{fig: uncoded_hist} (blue). 
In particular, out of $5000$ realizations, there are only $9$ MAN storage realizations worse than repetition placement realizations and $1621$ MAN placement realizations worse than cyclic placement realizations. 
Moreover, the MAN placement indeed achieves the minimum computing time in terms of both mean and variance compared to cyclic and repetition placements (see Table~\ref{tab: comparison}). 

\begin{table}[tb]
\centering
\caption{Comparison between MAN, cyclic and repetition placements.}
\small
\begin{tabular}{|c|c|c|c|}
\hline
computation time 
& cyclic & repetition & MAN  \\
\hline
mean  & $0.1492$  & $0.2296$ & $0.1442$ \\ \hline
variance  & $0.0033$ & 0.0114 & $0.0032$ \\
\hline
\end{tabular}
\label{tab: comparison} 
\end{table}

\begin{figure}
\centering
\includegraphics[width=0.8\columnwidth]{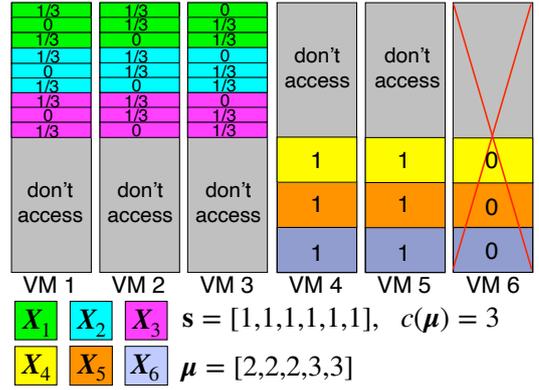} 
\caption{Illustration of uncoded USEC with straggler tolerance for $S=1$ using redundant task assignment.}
\label{fig: Uncoded_Elastic_Computing_Repetition_Straggler} 
\end{figure}

When the straggler tolerance is considered, we need to solve the optimization problem (\ref{eq: opt}) to obtain the optimal $\boldsymbol{M}^\star$ and then find a feasible computation assignment that meets $\boldsymbol{M}^\star$.  
Consider an example of a USEC system with homogeneous computing speed. Here, we let $N=N_t=6$, $J=3$, $S=1$, and the repetition placement is used. 
The optimal $\mu^\star[g,n], g \in [6], n \in [6]$ are shown in Fig.~\ref{fig: Uncoded_Elastic_Computing_Repetition_Straggler} and the optimal $\boldsymbol{\mu}^\star = [2,2,2,3,3]$. The optimal computation time is $c^\star(\boldsymbol{\mu}) = 3$.


\section{Proposed USEC Design} 

The proposed USEC design with straggler tolerance is given by Algorithm~\ref{algorithm:2}, which is obtained by 
solving the combinatorial optimization (\ref{eq: opt}) in a similar fashion as in \cite{woolsey2020heterogeneous} (line 6 in Algorithm~\ref{algorithm:2}). The proposed design is adaptive by measuring (line 14 in Algorithm~\ref{algorithm:2}) and updating (line 4 in Algorithm~\ref{algorithm:2}) the speed vector at time step. Interestingly, this algorithm adapts the previous CSEC (not USEC) computation assignment \cite{woolsey2020heterogeneous} to assign computations to $1+S$ machines. 

Next we will explain the proposed design. 
Since the proposed design without straggler tolerance is 
a special case of the general design with straggler tolerance for the 
combinatorial optimization problem (\ref{eq: opt}), 
then we will focus on designing algorithms to solve 
(\ref{eq: opt}). 

Similar to \cite{woolsey2020heterogeneous}, we will solve the combinatorial optimization problem (\ref{eq: opt}) exactly in two steps. In the first step, we solve the following relaxed convex optimization problem to obtain the optimal 
$\boldsymbol{M}^\star$
without considering whether such a computation assignment exists or not. 
\begin{subequations} \label{eq: uncoded opt 3}
\begin{align}
\underset{{\boldsymbol{M}} }{\text{minimize}} & \quad c\left(\boldsymbol{M}\right) =  \max_{n \in \Nc_t}  \frac{\sum_{g \in [G]}\mu[g,n]}{s[n]} \label{eq: uncoded 1} \\
\text{subject to:}  & \quad \sum_{n \in \Nc_t: \Xm_g \in \Zc_n} \mu[g,n] = 1+S, \forall g \in [G], \label{eq: usec constraint mu 3} \\
& \quad \mu[g,n] = 0, \forall \Xm_g \notin \Zc_n, n \in \Nc_t, \\
& \quad 0 \leq \mu[g,n] \leq 1, \forall n \in \Nc_t.
\end{align}
\end{subequations}
The difference between (\ref{eq: uncoded opt 3}) and (\ref{eq: uncoded opt 2}) is to change (\ref{eq: usec constraint mu 2}) from $\sum_{n \in  \Nc_t: \Xm_g \in \Zc_n} \mu[g,n] = 1, \forall g \in [G]$ to $\sum_{n \in  \Nc_t: \Xm_g \in \Zc_n} \mu[g,n] = 1+S, \forall g \in [G]$ as in (\ref{eq: usec constraint mu 3}). 
After obtaining the optimal $\boldsymbol{\Mm}^\star$, 
we will apply the {\em filling algorithm} developed in \cite{woolsey2020heterogeneous} to assign computations for each $\Xm_g \in \Zc_n, n \in \Nc_t$. 
Now we will describe the filling algorithm for USEC with homogeneous and heterogeneous computing speed, respectively. 

{\it Proposed USEC with homogeneous computation assignment}: Consider $\Nc_g = \{n: \Xm_g \in \Zc_n\}$ with $|\Nc_g| = N_g$. Then we define a computation assignment with $F_g=N_g$ row sets of $\Xm_g$. 
There are $N_g$ disjoint equally-sized row sets that collectively span all rows: $\mathcal{M}_{g,f}=\{1+(f-1)\frac{q}{N_gG},\ldots, f\frac{q}{N_gG} \}$ for $f\in[N_g]$. Then, define a cyclic assignment such that machine set $\mathcal{P}_{g,f}=\{f\%N_g,\ldots,(f+S)\%N_g\}$ for $f\in[N_g]$, where we define $a\%N_g \triangleq a-\left \lfloor \frac{a-1}{N_g} \right\rfloor N_g$ to facilitate the cyclic design.

{\it Proposed USEC with heterogeneous computation assignment}: 
Given the computation load matrix $\boldsymbol{M}^\star$, 
we can obtain the computation 
assignment by applying the assignment algorithm in \cite{woolsey2020heterogeneous} to assign computations to $1+S$ VMs for each $\Xm_g$ 
(line 6 in Algorithm~\ref{algorithm:2}). The computation assignment algorithm for $\Xm_g$ is given by Algorithm~\ref{algorithm:1}. 

%
%
%
%



\begin{remark}
  For both designs, we observe that the computation time $c(\boldsymbol{M})$ increases with the straggler tolerance, $S$. This demonstrates a trade-off between the computation time and straggler tolerance of the system.
\end{remark}



\begin{algorithm} 
  \caption{Adaptive Straggler Tolerant Uncoded Storage Elastic Computing}
  \label{algorithm:2}
  \begin{algorithmic}[1]
  \item[ {\bf Input}: $\hat{\boldsymbol{s}}$, $\gamma$, $S$, $T$, $\boldsymbol{w}_1$ ] 
   \hspace*{4cm} 
   \State $\boldsymbol{\nu}\leftarrow \hat{\boldsymbol{s}}$: same for all worker VMs
  \For {$t \in [T]$}
   \State {\bf At Master Machine}:
  \State \quad $\hat{\boldsymbol{s}}\leftarrow \gamma \boldsymbol{\nu} + (1-\gamma)\hat{\boldsymbol{s}}$  { (update estimate of speed vector)}.
  \State \quad $\mathcal{N}_t \leftarrow$ list of available machines 
  \State \quad 
  $\{F_g, \boldsymbol{\Mc}_g, \boldsymbol{\Pc}_g: \forall g \in [G]\} \leftarrow$ Results of computation assignment algorithm for $\Xm_g$ with straggler tolerance of $S$ for available machines $\mathcal{N}_t$ with speeds of $\hat{\boldsymbol{s}}$
  
  \State \quad Send $\boldsymbol{w}_t$ and 
  $\{F_g, \boldsymbol{\Mc}_g, \boldsymbol{\Pc}_g: \forall g \in [G]\}$ to worker VMs
   \State {\bf At Worker VMs}:
   \State \quad $n\leftarrow$ index of worker VM
   \State \quad $\mu[n] \leftarrow$ total computation load of worker VM $n$
  \State \quad $\tau_1\leftarrow$ current time
  \State \quad Perform assigned computations based on  $\{F_g, \boldsymbol{\Mc}_g, \boldsymbol{\Pc}_g: \forall g \in [G]\}$ 
  \State \quad $\tau_2\leftarrow$ current time
  \State \quad $\nu[n] \leftarrow \mu[n]/(\tau_2-\tau_1)$ {(calculate speed based on current time step)} 
  \State \quad Send computations and $\nu[n]$ to Master Machine
  \State {\bf At Master Machine}:
  {after receiving results from at most $N_t - S$ workers}.
  \State \quad $\boldsymbol{w}_{t+1}\leftarrow$ Combine worker results
  \EndFor
  \item[ {\bf Output}: $\boldsymbol{w}_T$ ]
  \end{algorithmic}
\end{algorithm}


\begin{algorithm}
  \caption{Computation Assignment for $\Xm_g$ for 
  Heterogeneous Computing Speed}
  \label{algorithm:1}
  \begin{algorithmic}[1]
  \item[ {\bf Input}: $\boldsymbol{\mu}_g^\star$, 
  $q$, $\boldsymbol{\Zc}$ and $\Nc_g = \{1,\cdots,N_g\}$. ]
  \item $\boldsymbol{m} \leftarrow \boldsymbol{\mu}_g^\star$
  \item $f \leftarrow 0$
  \While {$\boldsymbol{m}$ contains a non-zero element}
    \State $f \leftarrow f+1$
    \State $L' \leftarrow \sum_{i=1}^{N_g}m[i]$
    \State $N'\leftarrow$ number of non-zero elements in $\boldsymbol{m}$
    \State $\boldsymbol{\ell} \leftarrow$ indices that sort the non-zero elements of $\boldsymbol{m}$ from smallest to largest\footnotemark[5]
    \State $\mathcal{P}_{g,f} \leftarrow\{\ell [1], \ell [N'-L+2] , \ldots , \ell [N'] \}$
    \If {$N' \geq L+1$}
    \State $\alpha_{g,f} \leftarrow  \min \left(\frac{L'}{L} - m[\ell[N' - L + 1]], m[\ell[1]]\right)$\footnotemark[6] 
    \Else
    \State $\alpha_{g,f} \leftarrow  m[\ell[1]]$
    \EndIf
    \For {$n \in \mathcal{P}_{g,f}$}
    \State $m[n] \leftarrow m[n] - \alpha_{g,f}$
    \EndFor
  \EndWhile
  \item $F \leftarrow f$
  \State Partition rows $[\frac{q}{G}]$ of $\Xm_g$ into $F$ disjoint row sets 
  $\mathcal{M}_{g,1}, \ldots , \mathcal{M}_{g,F}$ of size $\frac{\alpha_1 q}{G},\ldots,\frac{\alpha_{F}q}{G}$ rows, respectively
  \item[ {\bf Output}: $F$, 
  $\{ \mathcal{M}_{g,1}, \ldots , \mathcal{M}_{g,F}\}$ and 
  $\{ \mathcal{P}_{g,1}, \ldots , \mathcal{P}_{g,F}\}$ ]
  \end{algorithmic}
\end{algorithm}
\footnotetext[5]{$\boldsymbol{\ell}$ is an $N'$-length vector and $0<m[\ell[1]]\leq m[\ell[2]]\leq \cdots \leq m[\ell[N']]$.}
\footnotetext[6]{This is the condition obtained by using Lemma~1 in \cite{woolsey2021cec}.}

\begin{figure}
\centering
\includegraphics[width=0.75\columnwidth]{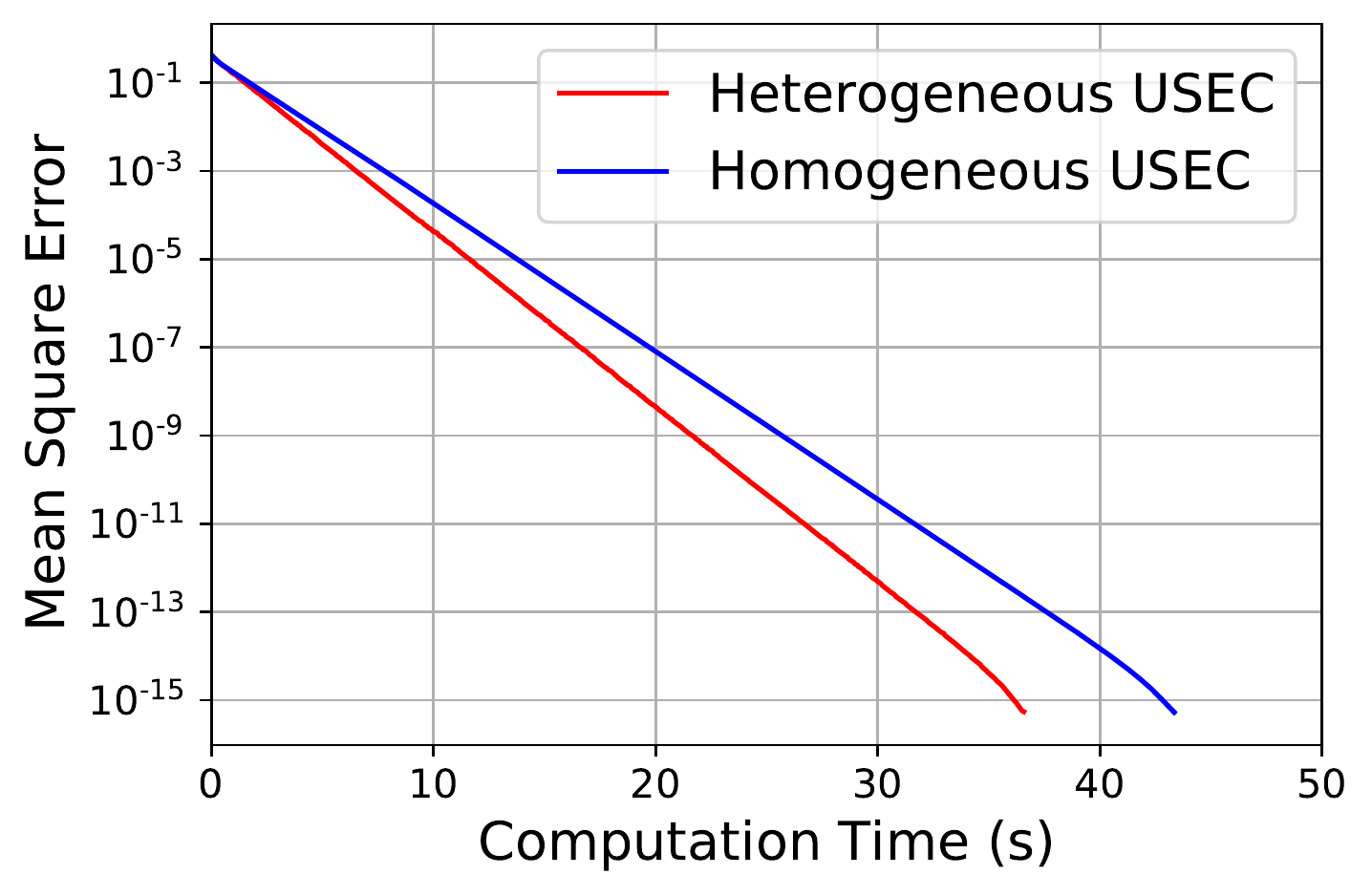}
\caption{{\bf Power Iteration}: Results using USEC designs on Amazon EC2 without stragglers (top) and with $2$ stragglers each iteration (bottom).  The y-axis represents the normalized mean square error between the true dominant eigenvector and the estimated eigenvector. 
}
\label{fig: Power}
\end{figure}

\section{Evaluations on Amazon EC2}


We evaluate the proposed algorithm using power iteration applications 
on Amazon EC2 instances. The goal is to compare the performance difference in terms of computation time between the homogeneous and heterogeneous task assignments. 

{\it Power Iteration}: 
The power iteration algorithm computes the largest eigenvalue and the corresponding eigenvector of a large matrix $\Xm$. In particular, it starts with a vector $\bv_0$, which may be an approximation to the dominant eigenvector or a random vector. The method is described by the recursive relation, $\bv_{k+1} = \frac{\Xm \bv_k}{\|\Xm \bv_k\|}$. The sequence $\bv_k$ converges to an eigenvector associated with the dominant eigenvalue. It can be seen that at each iteration, we can directly apply the proposed Algorithm~1. In particular, 
a dense $6,000$-by-$6,000$ symmetric matrix is row-wise split into $G=6$ sub-matrices which will be stored at each machine. 
We apply the repetition placement.
A vector of length $6,000$ is updated by performing a matrix-vector multiplication in a distributed manner on the available worker VMs. The master machine combines the results and normalizes the vector. This process is repeated such that the vector converges to the eigenvector associated with the largest eigenvalue.

The network has one \verb"t2.x2large" master machine with $8$ vCPUs and $32$ GiB of memory. The worker VMs consist of $3$ \verb"t2.large" instances, each with $2$ vCPUs and $8$ GiB of memory, and $3$ \verb"t2.xlarge" instances, each with $4$ vCPUs and $16$ GiB of memory. Similar to \cite{woolsey2021practicalcec}, we observed that all VMs have very different computing speed. For simplicity, we let $N=N_t$ and $S=0$ in order to show the advantage of the heterogeneous task assignment over the homogeneous task assignment. The result is shown in Fig.~\ref{fig: Power}, where the gain of Algorithm~\ref{algorithm:2} is about $20\%$ in terms of the computation time.

\section{Conclusions}
\label{sec: conclusions}
In this paper, we introduce a new optimization framework on USEC with heterogeneous computing speed to minimize the overall computation time. In particular, we consider the USEC systems under different uncoded storage placements and with or without straggler tolerance. For both scenarios, we propose optimal algorithms given the storage placements. 
These algorithms are evaluated using real applications on Amazon EC2 to demonstrate their gains in terms of computation time compared to the designs using the homogeneous computing speed assumption. 


\bibliographystyle{IEEEbib}
\bibliography{references_d2d}

\end{document}